\documentclass[a4paper,10pt]{article}
\usepackage{epsfig}
\usepackage{color}
\usepackage{amssymb}
\usepackage{amsmath}
\usepackage[ngerman,english]{babel}
\usepackage{multirow}
\usepackage{graphicx}
\usepackage{feynmf}
\usepackage[left=2.5cm,top=3cm,right=3cm,nohead]{geometry}

\newcommand{\etal}{{\it et al.}}
\newcommand{\pr}[4]{Phys. Rev. #1 {\bf #2}, #3 (#4)}
\newcommand{\hedp}[3]{High Energy Density Phys. {\bf #1}, #2 (#3)}
\newcommand{\astropj}[3]{Astrophys. J. {\bf #1}, #2 (#3)}

\newcommand{\cpp}[3]{Contrib. Plasmas Phys. {\bf #1}, #2 (#3)}
\newcommand{\physfluid}[3]{Phys. Fluids {\bf #1}, #2 (#3)}

\newcommand{\rmnumb}[2]{{#1}_{\rm #2}}
\newcommand{\rmupper}[3]{{#1}^{{\rm #2}}_{#3}}
\newcommand{\rmboth}[3]{{#1}^{{\rm #2}}_{\rm #3}}
\newcommand{\ket}[1]{| #1 \rangle}
\newcommand{\bra}[1]{\langle #1 |}

\newcommand{\bks}[1]{\left( #1 \right)}
\newcommand{\curlybra}[1]{\left\{ #1 \right\}}
\newcommand{\squarebra}[1]{\left[ #1 \right]}

\newcommand{\biggcurlybra}[1]{\bigg\{ #1 \bigg\}}
\newcommand{\Bigcurlybra}[1]{\Big\{ #1 \Big\}}

\newcommand{\impart}[0]{{\rm Im}\, }
\newcommand{\repart}[0]{{\rm Re}\, }

\newcommand{\absvalue}[1]{\left| {#1} \right| }

\newcommand{\kbt}[1]{\rmnumb{k}{B} T_{#1}}

\title{Ionization potential depression and optical spectra in a Debye plasma model}

\author{Chengliang Lin, Gerd R\"opke, Heidi Reinholz, and Wolf-Dietrich Kraeft}
\date{}
\begin{document}

\maketitle

\begin{abstract}
We show how optical spectra in dense plasmas are determined by the shift of energy levels as
well as the broadening owing to collisions with the plasma particles. In lowest approximation, 
the interaction with the plasma particles is described by the RPA dielectric function, leading to the 
Debye shift of the continuum edge. The bound states remain nearly un-shifted, their broadening 
is calculated in Born approximation. The role of ionization potential depression as well as 
the Inglis-Teller effect are shown. The model calculations have to be improved going beyond the 
lowest (RPA) approximation when applying to WDM spectra.
\end{abstract}

\section{Introduction}
Optical properties of dense plasmas have become of increasing interest with new experimental facilities exploring 
warm dense matter (WDM) 
and materials in the high-energy density  regime. Producing plasmas in the region of condensed matter densities and 
temperatures in the keV region, many-body effects such as the dissolution of bound states owing to the Mott effect
and the ionization potential depression (IPD) are within reach. New experiments
 \cite{Hoarty13,ciricosta12,Kraus16}  are devoted to explore these effects
which are essential for the ionization degree and the composition of high-energy density matter \cite{Chung05}.

Recently, there has been an intense discussion concerning bound state energies which are modified by the surrounding plasma.
In particular, experiments can not be described by traditional simple expressions for the IPD~\cite{EK63}.
A new quantum statistical approach has been worked out \cite{Lin17} which provides a better agreement with
experimental data. The shift of the continuum edge is determined by the single-particle self-energy (SE), 
and it has been shown that this is related to the dynamical structure factor.
In particular, the ion-ion structure factor is essential for the evaluation of the IPD.

Because the experimental observations are mainly related to the optical spectra, 
this is only an indirect observation of the IPD. Optical spectra are not only determined by the 
position of the energy levels of the radiator modified by the plasma surrounding, but also by the 
broadening produced by collisions. The broadening of spectral lines leads to the Inglis-Teller effect (ITE) \cite{ingtel39},
the peak structures of the optical spectra which are interpreted as spectral lines due to transitions
between bound states are washed out. If the overlap between neighbored lines is sufficiently large,
they are no longer visible as separate in the optical spectra.

For the interpretation of the disappearance of spectral lines at increasing plasma density there are two 
accepted but controversial scenarios. Firstly, the IPD becomes sufficiently large so that the bound state shifts
into the continuum and disappears.
Secondly, the ITE means that separate lines merge due to broadening so that peak structures are washed out.
We consider a more fundamental approach where many-particle effects for 
optical spectra are treated in a systematic way. Within this approach, both IPD and ITE are following from
the SE. The real part of the SE describes energy shifts (IPD), whereas its imaginary part is related to line 
broadening (ITE).

\section{Medium modifications}
The optical properties in a medium are determined via the absorption 
coefficient $\alpha(\omega)$, which is related to transversal part of the dielectric function (DF)
\begin{equation}
 n(\omega) + i c\, \alpha(\omega) /\omega = \lim_{q \rightarrow 0} \sqrt{\rmnumb{\epsilon}{tr} ({\mathbf{q}},\omega)} ,
\end{equation}
where $c$ is the speed of light in the vacuum and $n(\omega)$ denotes the refraction index. 
In the long wavelength limit, both the transversal and longitudinal part of the DF 
$\epsilon ({\mathbf{q}},\omega)$ are identical, i.e. 
$\rmnumb{\epsilon}{tr} ({\mathbf{q}},\omega) =\rmnumb{\epsilon}{long} ({\mathbf{q}},\omega)$. 
In addition, the longitudinal DF $\rmnumb{\epsilon}{long} ({\mathbf{q}},\omega)$
is determined by the polarization function 
$\Pi({\bf q}, \omega)$
\begin{equation}
 \rmnumb{\epsilon}{long} ({\mathbf{q}},\omega) = 1- \frac{e^2}{\varepsilon_0 q^2} \Pi({\mathbf{q}},\omega).
\end{equation}
The polarization function $\Pi ({\mathbf{q}},\omega)$ can be evaluated systematically 
using thermodynamic Green's functions and Feynman's diagrams. Subsequently, a systematic many-particle 
approach to the optical spectra of dense plasmas is given by the polarization function 
$\Pi({\bf q}, \omega)$ in the long wavelength limit. 
To obtain the line spectra, a cluster decomposition of the polarization function has to be performed \cite{KKER,RD},
\begin{equation}~\label{cluster}
 \Pi({\mathbf{q}},\omega)= \Pi_1({\mathbf{q}},\omega)+ \Pi_2({\mathbf{q}},\omega) + \cdots,
\end{equation}
which represent the one-particle $\Pi_1({\mathbf{q}},\omega)$ (single-particle states), 
two-particle contribution $\Pi_2({\mathbf{q}},\omega)$ (two-particle states) and so on.
Hence we consider the process $X^{Z+} \to X^{(Z+1)+} + e^- $ replacing the $Z$-fold charged ion (also including 
neutral atoms for $Z=0$) by a two-particle problem, the $(Z+1)$ core ion plus the electron.

We consider the interaction with the plasma in the lowest approximation,
i.e. the random phase approximation (RPA).
The two-particle propagator in $\Pi_2$ is modified by a two-particle SE.
The corresponding effective equation of motion is denoted as the Bethe-Salpeter equation which reads~\cite{Lin17,KKER,RKKKZ78}
\begin{align}
 \label{BSE}
 & \Big[E(1)+E(2)+ \sum_{\bf q}[f(1+{\bf q})+f(2-{\bf q})] V_{12}({\bf q}) 
  + \Delta V^{\rm eff}(1,2,{\bf q},z) \Big ] \! \psi(1,2,z) \nonumber \\
  + & \sum_{\bf q}\! \Big\{ [1\!-\!f(1)\!-\!f(2)] V_{12}({\bf q})
  + \Delta V^{\rm eff}(1,2,{\bf q},z)\Big\} \psi(1+{\bf q},2-{\bf q},z)\ =\ \hbar z \, \psi(1,2,z).
\end{align}
Here, the single particle states $1=\{ \hbar {\bf p}_1, \sigma_1,c_1\}$ are given by momentum, spin and species, 
respectively, $E(1) = \hbar^2 {\bf p}_1^2/(2 m_1)$. 
In the case considered here $c_1$ denotes the electron, charge number $Z_e=-1$, and $c_2$ the core ion. 
For the interaction we assume the Coulomb potential $V_{12}({\bf q}) = Z_{c_1}Z_{c_2} e^2/(\varepsilon_0 q^2)$.
$z$ is a complex variable following from the analytical continuation of 
the functions, defined for the Matsubara frequencies, into the entire $z$ plane. 
Of interest is the behavior of the functions near the real axis, $z=\omega \pm i \epsilon$.

The in-medium Schr\"odinger equation (\ref{BSE}) describes the influence of the medium by two effects, 
Pauli blocking and screening. Pauli blocking arises from the fact that
states which already are occupied by the medium are blocked. The blocking is described by 
the Fermi distribution function $ f(1)=[\exp(\beta (E(1)-\mu(1))+1]^{-1}$ in Eq.~\eqref{BSE}, with 
$\beta = 1/(k_B T)$ and $\mu(1)$ denoting the chemical potential of species 
$c_1$. 
Screening of the interaction by the medium is described by the effective interaction 
\begin{align}
 \label{Veff}
 \Delta V^{\rm eff}(1,2,{\bf q},z) & = -V_{12}({\bf q})\int_{-\infty}^\infty \frac{d \omega'}{\pi}
{\rm Im}\, \varepsilon^{-1}(q,\omega'+i0) \, \left[\rmnumb{n}{_B}(\omega')+1\right] \nonumber \\&
\times \left( \frac{\hbar}{\hbar z-\hbar \omega'-E(1)-E(2-{\bf q})}+\frac{\hbar}{\hbar z-\hbar \omega'-E(1+{\bf q})-E(2)}
\right) .
\end{align}
Including the effective potential, the effective Hamiltonian in the in-medium Schr\"odinger equation (\ref{BSE})
becomes complex and frequency dependent. This equation can be solved numerically~\cite{Seidel}, where both, real part and imaginary part
of the energy levels of the in-medium two-particle problem, can be calculated. In other words, 
the energy levels are shifted by the polarization of the medium and are also broadened by 
collisions with the plasma particles which leads to a finite life time of the energy levels. From Eq.~\eqref{BSE},
bound states are found at negative energies, whereas a continuum of scattering states is observed at positive energies with negative continuum edges. 

The remaining task is to determine the DF $\varepsilon(q,z)$  in the effective potential, i.e. Eq.~\eqref{Veff}.
A standard expression for the DF $\varepsilon(q,z)$ is the RPA DF. In the present work,
we will use the RPA DF to calculate the IPD and the spectral lines in plasmas. Within this simplest approximation,
we provide an insight to the explanation of the question how the ITE and the IPD can be systematically and consistently described 
within a unified theory.

\subsection{Shift of single particle states}
The definition of free states is generally described via $E_\mathbf{p} = \hbar^2\mathbf{p}^2/(2 m_c)$ for the atomic
species $c$ with the wavenumber $\mathbf{p}$. In a medium the interaction between particles must be considered.
Consequently, the definition of free states has to be modified by taking into account the interaction potential 
$V(\mathbf{r} \rightarrow \infty)$. Within the framework of Green's function technique, the shift of the continuum
edge is assumed to occur at ${\bf p}_1={\bf p}_2=0$ in Eq.~\eqref{Veff}.
Consequently, the influence of the surrounding plasmas
is described by~\cite{Lin17}
\begin{equation}\label{selfenergy}
  \rmupper{\Sigma}{corr}{c}(p,\omega) = \sum_{\mathbf{q}} \Delta V^{\rm eff}(p,-p,{\bf q},\omega) = -\sum_{\mathbf{q}}
 \int \frac{ d \omega'}{\pi} V_{cc}(q)\,  
 \frac{\curlybra{ 1+\rmnumb{n}{_B}(\omega')}\, \impart \varepsilon^{-1}(q,\omega'+i0)}{ \omega-\omega'-E_{c,{\bf p}+{\bf q}}/\hbar}.
\end{equation}
The shift of the continuum lowering is then given by the real part of this expression~\eqref{selfenergy}.
Following the theory in Ref.~\cite{Lin17}, the
imaginary part of the inverse DF in Eq.~\eqref{selfenergy}  can be
expressed via the dynamical structure factors according to the fluctuation-dissipation theorem. Considering only 
the ionic contribution to the continuum lowering, the shift of the continuum edge is obtained
\begin{equation}
 \rmupper{\Delta}{ion-ion}{c}(0,\omega) - {\cal P}\!\! \int \!\frac{d^3 {\bf q}}{(2\pi)^3}\!
 \int \!\frac{ d \omega'}{\pi} 
 \frac{V_{cc}(q)}{\omega-\omega'-E_{c,{\bf q}}/\hbar} 
  \frac{\pi Z_i e^2\, n_e}{\hbar \epsilon_0\, q^2} 
  \rmboth{S}{ZZ}{\rm ii}(q,\omega') ,
\end{equation}
where $ \rmboth{S}{ZZ}{\rm ii}(q,\omega')$ is the effective ion structure factor which includes
the screening cloud formed by the slowly moving electrons following the ionic motion~\cite{Lin17,GRHGR07}.
In the classical limit $(\hbar \omega = \hbar p^2 / (2 \rmnumb{m}{ion}) = 0)$ and plasmon pole approximation, 
the IPD is determined by the
difference between the SE before and that after the ionization of the investigated system, i.e.,
$ \rmboth{\Delta}{ion-ion}{_{IPD}} = \rmupper{\Delta}{ion-ion}{i}- (\rmupper{\Delta}{ion-ion}{e}
+\rmupper{\Delta}{ion-ion}{i+1})$. Then we have for the IPD
\begin{align}\label{final}
 \rmboth{\Delta}{ion-ion}{_{IPD}}=-\frac{(Z+1)e^2}{2 \pi^2 \epsilon_0 r_{_{\rm WS}} }\cdot 
 \frac{3 \Gamma_i}{ \sqrt{(9\pi/4)^{2/3} +3 \Gamma_i}} \cdot
 \int_0^\infty\frac{d q_0}{q_0^2}\rmboth{S}{ZZ}{\rm ii}(q_0),
\end{align}
where $q_0 = q/\bks{3\pi^2 n_i}^{1/3}$ is the reduced wavenumber. $\Gamma_i=Z^2 e^2/(4 \pi \epsilon_0 k_BT r_{_{\rm WS}})$ 
is the ionic coupling parameter with the Wigner-Seitz radius $r_{_{\rm WS}} =(4 \pi n_i/3)^{-1/3}$.

\subsection{Shift of two-particle states}
To investigate the optical spectra within the quantum statistical approach, we should return to the two-particle
problem, which is described by the atomic polarization function $\Pi_2$ in Eq.~\eqref{cluster}.
The pressure-broadened line profile is then given by~\cite{GHR91,OWR11}
\begin{align}
 {\cal L}(\omega)  = \sum_{ii'ff'} \frac{\omega^4}{8\pi^3\, c^3} \,e^{-\frac{\hbar \omega}{\kbt{}}}\,
 \bra{i} \mathbf{r} \ket{f} \bra{f'} \mathbf{r} \ket{i'} 
 \cdot\bra{i'}\bra{f'}\, \squarebra{\hbar \omega - \hbar \omega_{if} - \Sigma_{if}(\omega) + 
 i \rmupper{\Gamma}{v}{if} }^{-1}  \,\ket{i}\ket{f} 
\end{align}
with the SE for the initial $i$ and final $f$ states 
\begin{equation}
 \Sigma_{if}(\omega)  = \repart \curlybra{\Sigma_i(\omega) - \Sigma_f(\omega) } 
 + i \impart \curlybra{\Sigma_i(\omega) + \Sigma_f(\omega) }
\end{equation}
and the vertex correction
\begin{equation}
 \rmupper{\Gamma}{v}{if} = - \int \frac{d^3 \mathbf{p}}{(2\pi)^3}\int \frac{d^3 \mathbf{q}}{(2\pi)^3}
 \rmnumb{f}{e}(E_\mathbf{p} )\, V^2 (q)\,  M^0_{ii}({\mathbf q}) M^0_{ff}({-\mathbf q})\,
 \delta\bks{ \frac{\hbar^2 \mathbf{p\cdot q}}{\rmnumb{m}{e}} }
\end{equation}
where the unperturbed transition matrix elements are given by~\cite{OWR11}
\begin{equation}
 M^0_{n \alpha}({\mathbf q}) = i \curlybra{ Z \delta_{n\alpha} - \int d^3 \mathbf{r}\ 
 \psi_n^*( \mathbf{r})\, e^{i\mathbf{q\cdot r}}  \,\psi_\alpha( \mathbf{r}) } .
\end{equation}
Evidently, the shift of transition energy is given by the difference of real parts of the SE for
the initial $i$ and final $f$ states,
whereas the broadening of spectral lines is determined by the imaginary parts of the SE for
the initial $i$ and final $f$ states as well as the vertex correction $\rmupper{\Gamma}{v}{if}$.
Because of different dynamical time scales of ions and electrons, the electronic contribution is 
usually calculated in the impact approximation and the ionic contribution is described within the
quasi-static approximation~\cite{GHR91,Griem64}.

Introducing the microfield ansatz~\cite{GHR91}, the total SE $\Sigma_n(\omega,F)$ 
can be approximately decomposed into a frequency-independent ionic
part determined by the microfield strength $F$ and an only frequency-dependent electronic part. In this work,
we deal with the problem in a different way by treating the plasma electrons and ions at the same level, i.e. directly via the 
DF. The RPA DF for the non-degenerate case reads~\cite{KKER,Redmer97}
\begin{align}
 \rmnumb{\epsilon}{RPA}^{} ({\mathbf q},\omega) & = 
 1 + \frac{K}{2Q^3} \biggcurlybra{ 
 \sqrt{2} \squarebra{ D(x_1^+) - D(x_1^-)} + \sqrt{2/\gamma} \squarebra{ D(x_2^+) - D(x_2^-) } \nonumber \\
 - & i \Bigcurlybra{ \sqrt{\pi/2} \squarebra{ e^{-(x_1^{+})^2} - e^{-(x_1^{-})^2} } + 
 \sqrt{\pi/(2\gamma)} \squarebra{ e^{-(x_2^{+})^2} - e^{-(x_2^{-})^2} }
 }
  }
\end{align}
with the abbreviations
\begin{align}
 & Q = \frac{\hbar q}{\bks{\rmnumb{m}{e}\, \kbt{}}^{1/2}}\, , \qquad 
 K = \frac{(1+z^2)\hbar^2 \rmnumb{n}{e} e^2 }{\rmnumb{m}{e}\,\varepsilon_0\, (\kbt{})^2 }\, , \qquad 
 \gamma = \frac{\rmnumb{m}{e}}{\rmnumb{m}{ion}}\,  \nonumber \\
 & x_1^{\pm} = \sqrt{\frac{1}{2}}\, \squarebra{ \frac{\omega}{q}\bks{\frac{\rmnumb{m}{e}}{\kbt{}}} \pm \frac{Q}{2}  }\, , \qquad 
 x_2^{\pm} = \sqrt{\frac{1}{2\gamma}}\, \squarebra{ \frac{\omega}{q}\bks{\frac{\rmnumb{m}{e}}{\kbt{}}} \pm \gamma \frac{Q}{2}  },
\end{align}
where $D(x) = \exp\bks{-x^2} \int_0^x dt\, \exp\bks{-t^2}$ is the Dawson integral. Inserting this expression into the 
Eq.~\eqref{selfenergy},
an expression for the IPD can be obtained, which is the Debye shift for the IPD in the low density case and is equivalent 
to the expression~\eqref{final}. Instead of starting from Eq.~\eqref{BSE}, we calculate the SE of bound two-particle states 
with the following expression~\cite{RD,GHR91}
\begin{align}~\label{sfRPA}
\rmupper{\Sigma}{}{n}  (\omega) =  \int \frac{d^3 {\mathbf q}}{(2 \pi \hbar)^3} V(q) \,
 \sum_{\alpha} \absvalue{M^0_{n \alpha}({\mathbf q})}^2 
 \int_{-\infty}^{\infty} \frac{d \omega}{\pi}
 \frac{\squarebra{1 + \rmnumb{n}{_B}(\omega) } \cdot \impart \rmnumb{\epsilon}{RPA}^{-1} ({\mathbf q},\omega + i0) }
 {E_{n} - E_{\alpha} - \bks{ \hbar \omega + i 0} },
\end{align}
which is the SE in dynamically screened Born approximation with the RPA DF.
Accounting for the thermal motion of the plasma ions, which results in the Doppler broadening, the full line profile
is given by a convolution~\cite{Griem64}
\begin{equation}
 \rmupper{I}{total}{}(\Delta \omega) = \frac{c}{\omega_0} \sqrt{\frac{\rmnumb{m}{ion}}{2\pi \kbt{} }} 
 \int_{-\infty}^{\infty} d \Delta \omega' \cdot  \rmupper{I}{pr}{}(\Delta \omega') \cdot 
 \exp\curlybra{ - \frac{\rmnumb{m}{ion} c^2 }{2\kbt{}}\, 
 \bks{ \frac{ \Delta \omega - \Delta \omega' }{ \omega_0 + \Delta \omega' } }^2 } .
\end{equation}

\section{Results and discussion}

\begin{figure}[ht]
\centering 
\includegraphics[width=0.60\textwidth]{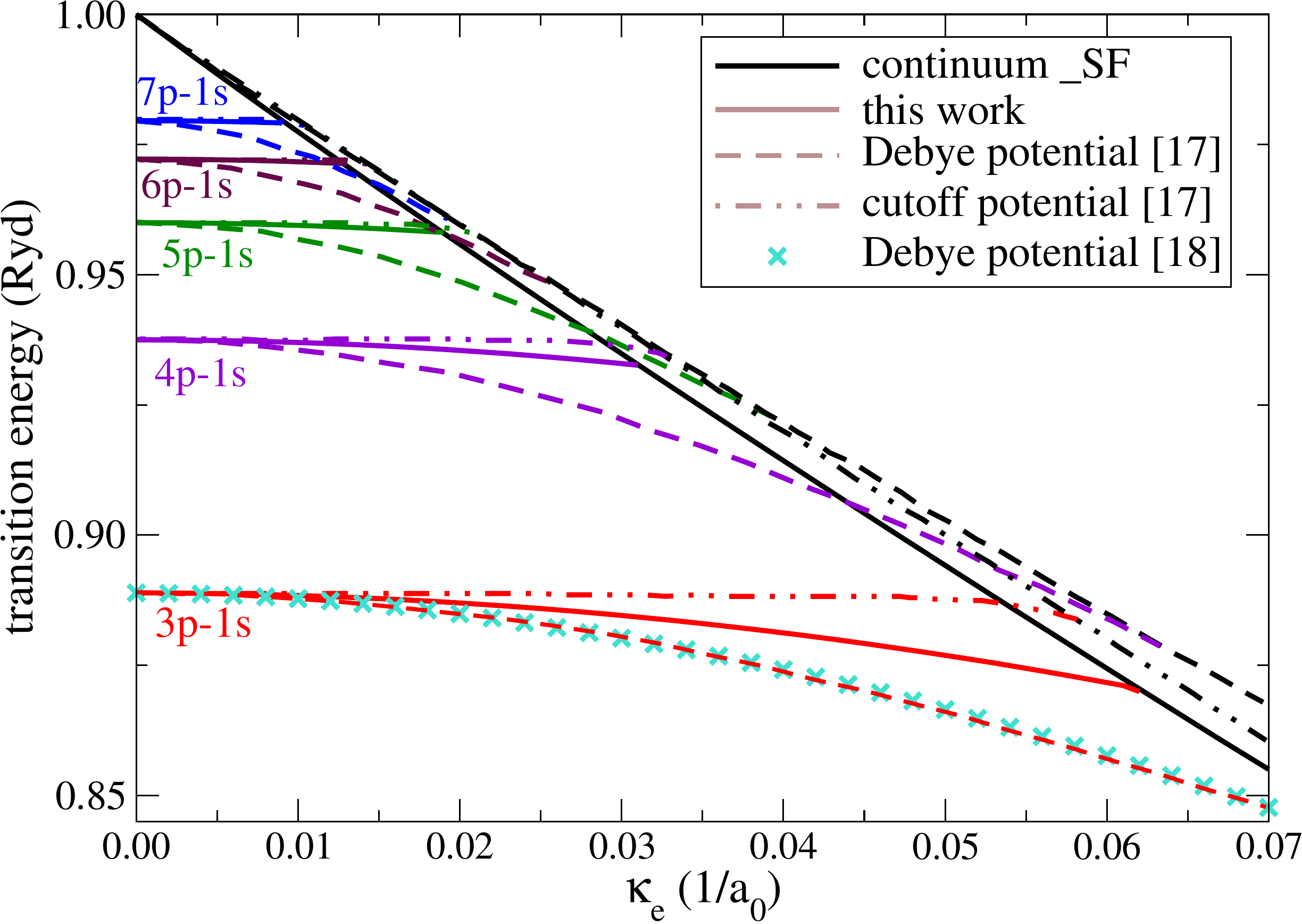}
\caption{Transition energy for the Lyman series $\hbar \omega = E_{n\mathrm{p}} - \rmnumb{E}{1s}$ and the 
threshold energy for the bound-free transition $\hbar \rmnumb{\omega}{th} = V(\infty) - \rmnumb{E}{1s}$
with respect to the inverse Debye length of electrons $\rmnumb{\kappa}{e} $. 
The solid lines represent the results of this work. The dashed lines and the dotted-dashed lines describe the 
numerical results of the Schr\"odinger equation with a Debye potential~\cite{Hoe82,Qi09} and a cutoff potential~\cite{Hoe82}, 
respectively. The cross points are also calculated with Debye potential~\cite{Qi09}.}
\label{shiftResults}
\end{figure}

\begin{figure}[h]
\centering 
\includegraphics[width=0.60\textwidth]{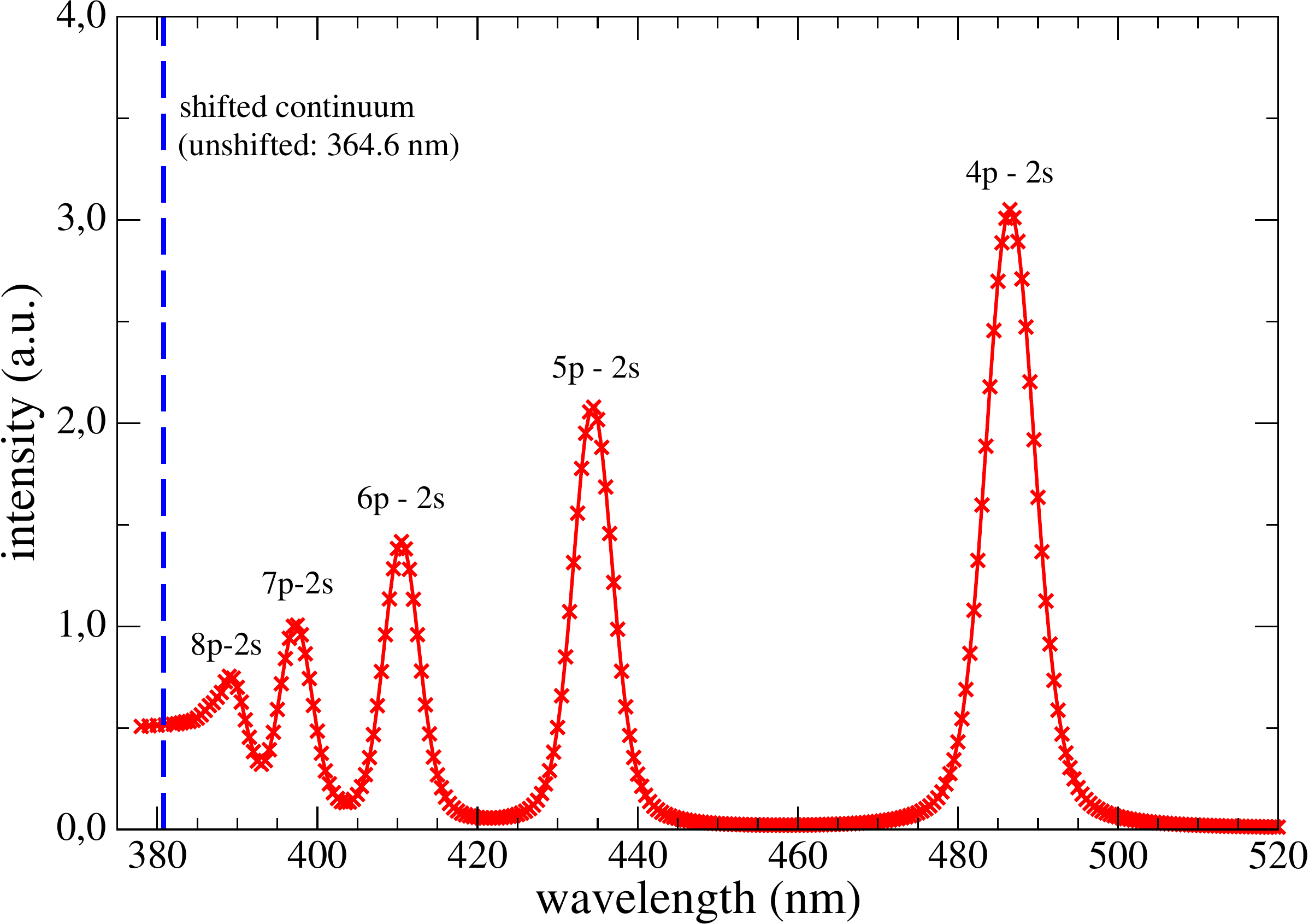}
\caption{Optical spectra of a plasma at the electron density $ \rmnumb{n}{e} = 2.1\cdot 10^{17}\, \mathrm{cm}^{-3}$ 
and temperature $ \rmnumb{T}{e} = 1.02$ eV calculated within RPA~\eqref{sfRPA}. Gaussian broadening with FWHM of 0.002 Ryd is considered in the calculation.
Both ITE and IPD are systematically calculated from the quantum statistical approach. A continuum background~\cite{OWR11}
is considered.}
\label{lineShapeResults}
\end{figure}

The interaction of a radiating particles with the plasma environment results in both shift and broadening of
the eigenlevels of a two-particle state. At first, we concentrate on the shifts of both the bound states and the free states.
In the present work, we calculate the IPD and the shift of bound states via Eq.~\eqref{final} and Eq.~\eqref{sfRPA},
respectively. Comparison of our results for the transition
energies and bound-free threshold energies
and the results predicted by other theoretical approaches is shown is Fig.~\ref{shiftResults}.

The positions of emission line centers are determined by the transition energies.
One of the commonly used theoretical approaches is to solve the Schr\"odinger equation with a 
pseudo-potential~\cite{Hoe82,Qi09} for eigenenergies and the corresponding wave functions. Numerical solutions have been obtained
using the Debye potential~\cite{Hoe82,Qi09} or the cutoff potential~\cite{Hoe82,RG86} as the pseudo-potential.
From Fig.~\ref{shiftResults}, it is seen that the line center positions derived from the cutoff potentials are almost unchanged
in comparison to the pure Coulomb case. In contrast, the Debye potential predicts a strong modification
for the transition energy, in particular for relatively high densities
($\rmnumb{\kappa}{e} = \sqrt{\rmnumb{n}{e} e^2/(\varepsilon_0 \kbt{}) } \geq 0.02 /a_0$).
However, this result is in strong contradiction to the experimental measurements, where the positions of line centers
are only slightly shifted~\cite{RG86}. 
In the Debye model, the shifts do not separately depend on density and temperature but only
on the screening parameter $\rmnumb{\kappa}{e}$.

The optical spectra of the Balmer series calculated within RPA combined with a continuum background due 
to Bremsstrahlung are displayed in Fig.~\ref{lineShapeResults}.
For the given plasma conditions ($ \rmnumb{n}{e} = 2.1\cdot 10^{17}\, \mathrm{cm}^{-3}$, $ \rmnumb{T}{e} = 1.02$ eV),
the IPD value is about $0.138$ eV, which means that the energy levels appear only up to $n=9$ and
the Balmer limit is shifted from $364.6$ nm to $380.8$ nm. Another feature of the predicted line shape (i.e. ITE) is the 
formation of quasi-continuum states, i.e. the band structure, starting from the energy level $n=6$.
Moreover, the transition peak from $\mathrm{9p} \rightarrow \mathrm{2s} $ is washed out due to the broadening. 
In summary, to calculate the spectra in the whole wavelength range, the first step is to determine the highest existing
quantum state from the IPD theory. The second step is to calculate the line shape up to the highest energy level predicted
by the IPD theory. Actually, the optical spectra of Balmer series under these conditions were experimentally measured by
Goto \etal~\cite{Goto07}, where the disappearance of the transition peaks is observed already starting from a lower principal 
quantum number $n=6$. The reason
for this disagreement arises from the fact that the ionic contribution to the broadening is excessively underestimated by RPA.
Comparing the experimentally measured line broadening of H$_\alpha$ line with the RPA result, it is shown that
the RPA result is only $25\%$ of the experimental value.

The line merging due to the broadening does not have a simple relation to the shift of the continuum edge.
Generally, both effects have to be considered carefully and comprehensively for correctly reproducing
the experimental spectra. For weakly coupled and non-degenerate plasmas, 
the line merging occurs at lower principal quantum numbers in comparison to 
the IPD effect as shown in Fig.~\ref{lineShapeResults} and measured in laser produced plasmas~\cite{Hoarty13,Nantel98}. 
Recently, the optical spectra in highly charged aluminum plasmas are observed, where the 
continuum lowering is so large that the lines disappear before they are subject to
broadening sufficient to merge them~\cite{Hoarty13}. 
Therefore,
it is of relevance to develop an
accurate spectral line theory using the systematic quantum statistical approach, 
since spectroscopic methods are the most reliable tool to analyze
density and temperature conditions.

\section{Conclusions, further improvements}

We have shown that both effects, IPD and ITE, have to be considered to explain the disappearance of spectral lines.
The synthetic spectra and the transition energies are systematically described using a consistent quantum statistical
approach within the simplest RPA. The Debye results for both IPD and optical transitions are obtained.
Of course, the Debye model is very simple and the ionic contributions are strongly underestimated. 
For the hydrogen plasmas, the ion contribution is more important. The RPA is not satisfied and has to be improved.
The underestimation might be rendered by using the ionic structure factor to calculate the line broadening and shift like in the 
case for the IPD when going beyond the RPA~\cite{Lin17}. Additionally,
the electron contribution can be improved by introducing the T matrix 
describing strong collisions. Furthermore, photon re-absorption plays an important role for the transport of radiation in a plasma,
which has a strong influence on the transitions from the low quantum numbers~\cite{OWR11}. This has to be taken into account in
calculating the line shapes. For higher levels, the transition rates may be better described via semi-classical description, 
for example using wave-packet states for the electrons~\cite{Lin16}.

In the context of new experimental facilities exploring warm dense matter~\cite{Hoarty13,ciricosta12,Kraus16,Nantel98}, strongly 
coupled and nearly degenerate Coulomb systems can be produced. 
A detailed description of the spectrum emitted from such systems in equilibrium and non-equilibrium conditions is still a challenging problem.
As shown in this work, the quantum statistical approach based on thermodynamic Green's function technique provides a possibility
to understand the many-body systems under such extreme conditions.

\section*{Acknowledgement}
 This work is supported by the German Research Foundation DFG within SFB 652.


\begin{thebibliography}{10}



 \bibitem{Hoarty13}
 D. J. Hoarty \etal, 
 \pr{Lett.}{110}{265003}{2013}.
 
  \bibitem{ciricosta12}
 O. Ciricosta \etal, 
 \pr{Lett.}{109}{065002}{2012}; Nat. Commun. {\bf 7}, 11713 (2016).
 

\bibitem{Kraus16}
D. Kraus \etal,
Phys. Rev. E {\bf 94}, 011202( R) (2016).
%

 \bibitem{Chung05}
H.-K. Chung \etal, High Energ. Dens. Phys. {\bf 1}, 3 (2005).

\bibitem{EK63}
 G. Ecker and W. Kr\"oll, 
\physfluid{6}{62}{1963}; 
 J. C. Stewart and K. D. Pyatt, Jr., 
 \astropj{144}{1203}{1966}.

\bibitem{Lin17}
C. Lin \etal., \pr{E}{96}{013202}{2017}.

\bibitem{ingtel39}
D. R. Inglis and E. Teller,
ApJ {\bf 90}, 439, (1939).



 
%
%
%
%
%
%
%
%
%


   \bibitem{KKER}
W.-D. Kraeft \etal, {\it Quantum Statistics of Charged Particle Systems} 
(Akademie-Verlag Berlin, 1986).


\bibitem{RD}
G. R\"opke and R. Der, 
 phys. stat. sol. (b) {\bf 92}, 501 (1979).

  \bibitem{RKKKZ78}
 R. Zimmermann \etal, 
 phys. stat. sol. (b) {\bf 88}, K59 (1978); phys. stat. sol. (b) {\bf 90}, 175 (1978).

\bibitem{Seidel}
J. Seidel, S. Arndt, and W.-D. Kraeft, Phys. Rev. E {\bf 52}, 5387 (1995).



\bibitem{GRHGR07}
G. Gregori, A. Ravasio, A. H\"oll, S. H. Glenzer and S. J. Rose, 
\hedp{3}{99}{2007}.

 \bibitem{GHR91}
S. G\"unter, L. Hitzschke, and G. R\"opke, Phys. Rev. A {\bf{44}}, 6834 (1991).

\bibitem{OWR11}
B. Omar, A. Wierling, and G. R\"opke,
\cpp{51}{22}{2011}.


 
%
%
%

  


%
 

%



\bibitem{Griem64}
H. Griem, {\it Plasma Spectroscopy}, ( McGraw-Hill 1964).



\bibitem{Redmer97}
R. Redmer,
Physics Reports \textbf{282}, 35 (1997).




\bibitem{Hoe82}
F. E. H\"ohne and R. Zimmermann, J. Phys. B: At. Mol. Phys. {\bf 15}, 25551 (1985).

\bibitem{Qi09}
Y. Y. Qi \etal, Physics of Plasmas \textbf{16}, 023502 (2009).

\bibitem{RG86}
R. Radtke and K. G\"unther, 
\cpp{26}{143}{1986}; \cpp{26}{151}{1986}.




\bibitem{Goto07}
M Goto \etal,
Plasma Phys. Control. Fusion \textbf{49}, 1163 (2007).






\bibitem{Nantel98}
M. Nantel \etal, \pr{Lett.}{80}{4442}{1998}.

\bibitem{Lin16}
C. Lin \etal, \pr{E}{93}{042711}{2016}.

\end{thebibliography}
\end{document}